\documentclass[fleqn,usenatbib]{mnras}

\usepackage[svgnames]{xcolor}
\usepackage{newtxtext,newtxmath} 
\usepackage{graphicx}          
\usepackage{microtype}

\usepackage{algorithm}
\usepackage[commentColor=LimeGreen]{algpseudocodex}

\graphicspath{{./figures/}}
\hypersetup{linkcolor=red}     

\algrenewcommand{\algorithmicrequire}{\textbf{Input:}}
\algrenewcommand{\algorithmicensure}{\textbf{Output:}}

\newcommand{\kms}{\text{km s}^{-1}}
\newcommand{\dd}{\mathrm{d}}
\newcommand{\ppxf}{\textsc{ppxf}}
\newcommand{\norm}[1]{\lVert#1\rVert}
\newcommand{\sn}{\mathcal{S/N}}
\newcommand{\snb}{\mathcal{S/N}_\mathrm{bin}}
\newcommand{\snt}{\mathcal{S/N}_\mathrm{T}}

\def\equationautorefname~#1\null{equation~(#1)\null}

\title[PowerBin: Data Binning]{PowerBin: Fast Adaptive Data Binning with Centroidal Power Diagrams}

\author[M.~Cappellari]{Michele Cappellari\thanks{E-mail:
michele.cappellari@physics.ox.ac.uk}\\
Sub-Department of Astrophysics, Department of Physics, University of Oxford, Denys Wilkinson Building, Keble Road, Oxford, OX1 3RH, UK}

\date{Accepted 2025 October 06. Received 2025 October 06; in original form 2025 September 08}

\pagerange{\pageref{firstpage}--\pageref{lastpage}} \pubyear{2025}

\begin{document}

\label{firstpage}
\maketitle

\begin{abstract}
    Adaptive binning is a crucial step in the analysis of large astronomical datasets, such as those from integral-field spectroscopy, to ensure a sufficient signal-to-noise ratio ($\sn$) for reliable model fitting. However, the widely-used Voronoi-binning method and its variants suffer from two key limitations: they scale poorly with data size, often as $\mathcal{O}(N^2)$, creating a computational bottleneck for modern surveys, and they can produce undesirable non-convex or disconnected bins. I introduce \textsc{PowerBin}, a new algorithm that overcomes these issues. I frame the binning problem within the theory of optimal transport, for which the solution is a Centroidal Power Diagram (CPD), guaranteeing convex bins. Instead of formal CPD solvers, which are unstable with real data, I develop a fast and robust heuristic based on a physical analogy of packed soap bubbles. This method reliably enforces capacity constraints even for non-additive measures like $\sn$ with correlated noise. I also present a new bin-accretion algorithm with $\mathcal{O}(N\log N)$ complexity, removing the previous bottleneck. The combined \textsc{PowerBin} algorithm scales as $\mathcal{O}(N\log N)$, making it about two orders of magnitude faster than previous methods on million-pixel datasets. I demonstrate its performance on a range of simulated and real data, showing it produces high-quality, convex tessellations with excellent $\sn$ uniformity. The public Python implementation provides a fast, robust, and scalable tool for the analysis of modern astronomical data.
\end{abstract}


\section{Introduction}
\label{sec:intro}

\begin{figure*}
        \includegraphics[width=\textwidth]{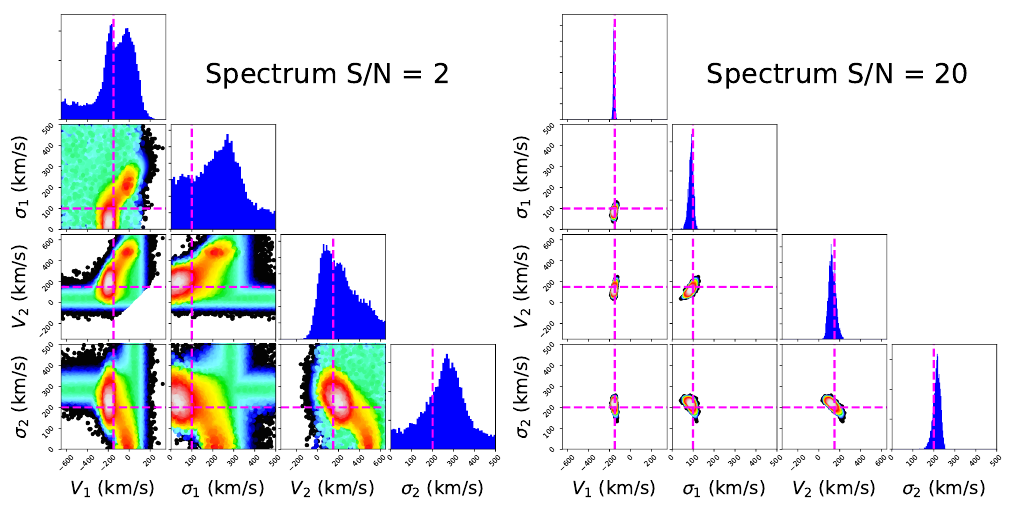}
        \caption{Posterior probability distributions for the four kinematic parameters
            $(V_1,\sigma_1,V_2,\sigma_2)$ of two stellar components, recovered
            with \ppxf\ coupled to an Adaptive Metropolis MCMC sampler ($10^5$ steps).
            Left: spectrum with $\sn=2$ per pixel. The joint and marginal
            posteriors are highly non-Gaussian, multimodal and biased away from
            the true input values (magenta dashed lines), illustrating that
            parameter estimates at low $\sn$ are unreliable. Right: spectrum with
            $\sn=20$ per pixel. The posteriors become well behaved, approximately
            Gaussian and centred on the true values, allowing robust inference.
            Contours mark the 68 and 95 per cent credible regions; diagonal
            panels show marginal histograms. This comparison demonstrates why
            one must bin spaxels to reach sufficient $\sn$ before fitting
            non-linear spectral models.
        \label{fig:corner_plot}}
\end{figure*}

Modern astronomical surveys, particularly those using integral-field spectroscopy (IFS), produce vast, spatially-resolved datasets containing millions of spectra \citep[e.g.,][]{Cappellari2011p1, Sanchez2012, Bryant2015, Bundy2015}. A common task is to fit these data with complex physical models to extract quantities like stellar kinematics \citep[e.g.][]{Westfall2019}, star formation histories, or chemical abundances \citep[e.g.,][]{McDermid2015, Scott2017, Lu2023}. However, the signal-to-noise ratio $\sn$ of individual spatial pixels (spaxels) is often too low for reliable model fitting \citep{Cappellari2003}. Fitting a non-linear model to low-$\sn$ data typically yields a highly non-Gaussian posterior probability distribution for the model parameters, making it difficult to derive meaningful best-fitting values and uncertainties from Bayesian methods \citep[e.g.][]{gelman2014bayes}. Crucially, simply averaging the biased results from low-$\sn$ fits does not recover the true parameters, because the posterior does not need to be symmetric around the true parameters.

The most effective solution is to combine the data from adjacent spaxels into larger bins to reach a sufficient $\sn$ \textit{before} performing the model fit. This process, known as adaptive binning, is a crucial preprocessing step in the analysis of IFS data and other 2D datasets like X-ray images \citep[e.g.,][]{Sanders2004, Diehl2007}.

\subsection{The Voronoi-Binning Method}

To address this need, \citet{Cappellari2003} introduced the Voronoi-binning algorithm, implemented, e.g., in the \textsc{VorBin} package\footnote{Python version 3.1 from \url{https://pypi.org/project/vorbin/}}. This method has become a standard tool in astrophysics for partitioning 2D data to meet three key criteria for an optimal binning scheme:  
(i) the bins must form a complete, non-overlapping tessellation of the data;  
(ii) the bins should be as compact (round) as possible to preserve spatial resolution; and  
(iii) a user-defined scalar function of each bin should be as uniform as possible around a target value.  

This optimized function is entirely general: while it is often chosen to be the bin’s $\sn$, the algorithm places no restriction on its form. It may represent, for example, the fractional error in a physical parameter derived from a spectral fit to the bin’s data, or a composite metric such as a weighted combination of $\sn$ values measured in different photometric bands over the same bin area.  

The algorithm achieved this through a two-stage process \citep{Cappellari2003}. First, a \textit{bin-accretion stage} provides an initial tessellation satisfying the target $\sn$. This greedy algorithm starts with the unbinned pixel with the highest $\sn$, accretes its nearest neighbours until the target $\sn$ is reached, and repeats this process until all pixels are binned. Second, an iterative \textit{regularization stage} improves the bin morphology using a Centroidal Voronoi Tessellation \citep[CVT,][]{Du1999}. In each iteration, a Voronoi tessellation is generated from the current bin generators, and these generators are then updated to be the new centroids of their corresponding cells, making the bins more compact.

\subsection{Limitations of Existing Methods}

Despite its widespread success, the evolution of astronomical surveys has revealed some limitations of the original method and its subsequent extensions.
 
\begin{itemize}
\item \textbf{Non-Convexity:} An important extension by \citet{Diehl2006}, which is included in \textsc{VorBin}, introduced a \emph{multiplicatively-weighted} Voronoi tessellation to improve $\sn$ uniformity and bin shapes. While effective, the adopted tessellation sacrifices a key morphological property: the guarantee of convex bins of the original CVT method. Although not a common occurrence, this can result in undesirable non-convex or even disconnected bins, which can be problematic in certain situations.

\item \textbf{Computational Speed:} The original implementation was not designed for the massive and rapidly growing datasets of modern and future astronomical surveys. Instruments like MUSE \citep{Bacon2010}, employed in large surveys \citep[e.g.][]{Sarzi2018,Gadotti2019,Emsellem2022,FraserMcKelvie2025}, can already produce mosaics of millions of pixels---for example, the 3 and 9 million spaxel mosaics of M42 and NGC~253 by \citet{Weilbacher2015} and \citet{Congiu2025} respectively---and this trend is set to accelerate. Next-generation concepts like the Wide-field Spectroscopic Telescope (WST) \citep{Bacon2024} are projected to produce datasets orders of magnitude larger still. Both the bin-accretion and the iterative regularization stages of \textsc{VorBin} scale poorly with the number of input spaxels, creating a significant computational bottleneck. In particular, the multiplicatively-weighted Voronoi diagram has a fundamental time complexity of $\mathcal{O}(n^2)$, where $n$ is the number of bins \citep{Aurenhammer1984}, making it impractical for large $n$, and the same is true for the bin-accretion algorithm.

\end{itemize}  

The goal of this paper is to introduce \textsc{PowerBin}, a fast and robust algorithm that addresses the limitations of previous adaptive-binning methods. I recast adaptive binning as a semi-discrete optimal-transport, or data-quantization, problem whose solutions are Centroidal Power Diagrams \citep[CPDs,][]{Aurenhammer1987,Aurenhammer1998,Merigot2011,DeGoes2012,Levy2015}. Building on a simple geometric/physical insight, \textsc{PowerBin} iteratively adjusts the power-diagram weights to enforce per-bin capacity targets while keeping cells convex and compact. The resulting scheme is computationally efficient, stable in the presence of realistic, non-additive capacity measures (for example, when pixel noise is correlated), and scales to the large datasets produced by modern astronomical surveys. A public Python reference implementation accompanies this paper.

This paper is structured as follows. \autoref{sec:examples} provides practical examples illustrating the necessity of binning. \autoref{sec:voronoi_generalizations} reviews the family of weighted Voronoi diagrams. \autoref{sec:cpd_ot} introduces the optimal transport framework and its connection to Centroidal Power Diagrams. \autoref{sec:fast_power_insight} presents the core physical insight behind my new fast regularization algorithm. \autoref{sec:accretion} describes the new fast bin-accretion algorithm. \autoref{sec:applications} demonstrates the performance of the new method on real and simulated data. \autoref{sec:benchmark} presents execution-time benchmarks. Finally, \autoref{sec:conclusions} summarizes my findings.

\section{Examples Illustrating the Need for Binning}
\label{sec:examples}

\begin{figure*}
    \includegraphics[width=.8\textwidth]{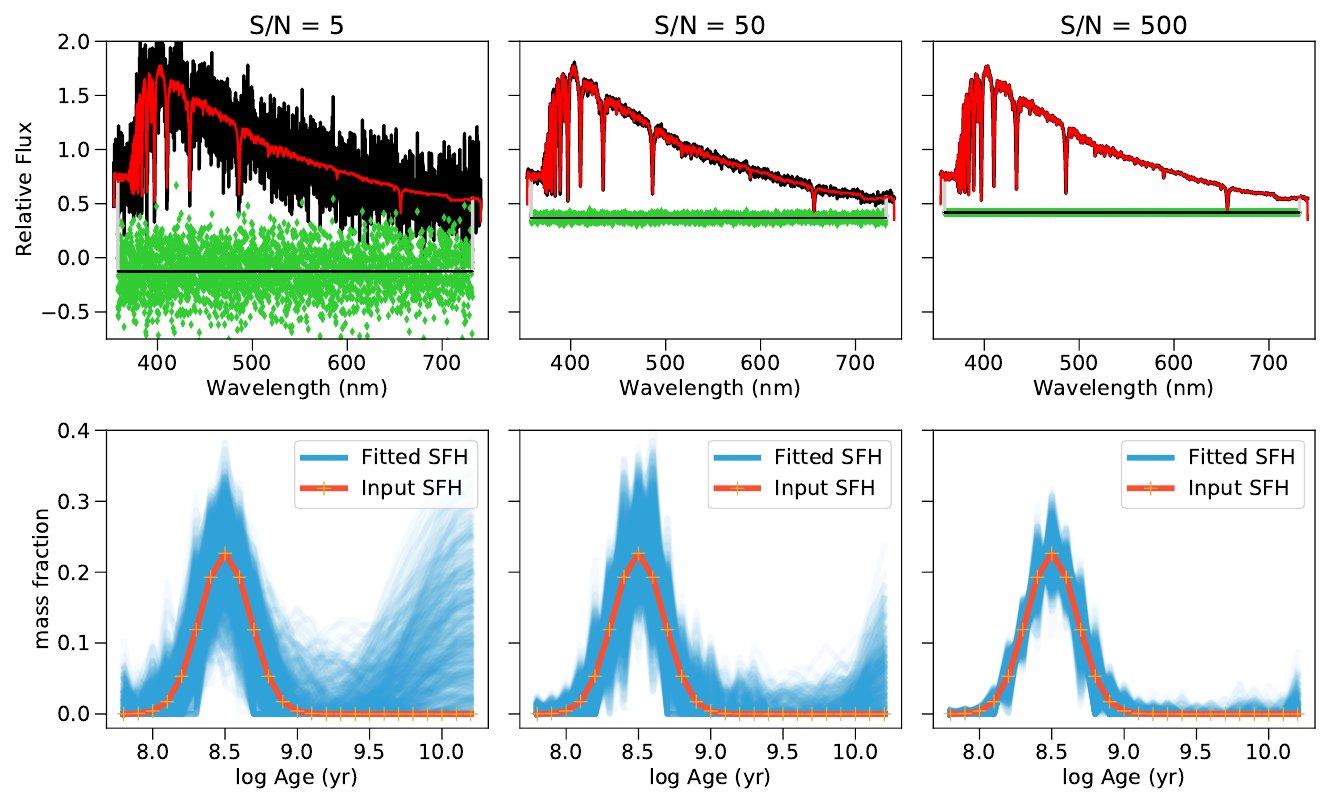}
        \caption{Recovery of a simple input star-formation history from full-spectrum fitting at three signal-to-noise levels. Top row: mock spectrum (black), \ppxf\ best-fit model (red) and residuals (green) for $\sn = 5$, 50 and 500 (left to right). Bottom row: the distribution of 1000 Monte Carlo recovered SFHs (transparent blue lines) compared to the input single burst at 0.3\,Gyr (orange line with markers). At low $\sn$ the recovered SFHs are biased and highly scattered; increasing $\sn$ progressively reduces bias and scatter, and by $\sn=500$ the input burst is recovered with high fidelity. The fits employ 25 solar-metallicity MILES templates and 1000 Monte Carlo realizations per $\sn$ level.
        \label{fig:ppxf_sfh}}
\end{figure*}

While \autoref{sec:intro} described the general motivation for binning, the importance of this preprocessing step is best understood through practical examples. The core issue is that fitting complex, non-linear models to low-$\sn$ data can lead to results that are not just uncertain, but systematically biased. This section presents two common scenarios in spectral analysis that demonstrate this effect and motivate the need for an optimal binning strategy. The behaviour shown is generic and applies to many types of data analysis in other fields. An ideal binning scheme should partition the data according to three criteria \citep{Cappellari2003}: (1) a topological criterion, ensuring all data are used without overlap; (2) a morphological criterion, requiring bins to be as compact (round) as possible to preserve spatial resolution; and (3) a uniformity criterion, minimizing the $\sn$ scatter around a target value. The following examples highlight why achieving a target $\sn$ is paramount.

\subsection{Extracting Stellar Kinematics of Multiple Components}

This example examines the problem of separating the kinematics of multiple stellar populations along a single line of sight. The goal is to illustrate a generic characteristic of non-linear problems in general, and astrophysical spectrum fitting in particular: at low $\sn$ the model posterior is not informative of the true solution. For this reason, the choice of spectra and kinematics is rather arbitrary and is not meant to represent a specific situation. I construct a synthetic spectrum made from two distinct stellar components and attempt to recover their velocities and dispersions at two representative per-pixel signal-to-noise levels, showing that low $\sn$ yields multimodal, biased parameters while higher $\sn$ permits reliable parameter recovery.

For this test, I took two model spectra from the MILES library \citep{Vazdekis2010}, both with solar metallicity but with different ages (12.6 Gyr and 1.0 Gyr). The spectra were logarithmically rebinned to a velocity scale of 70 $\kms$, typical of large surveys like SDSS \citep[e.g.][]{Abdurrouf2022}, and normalized to contribute equally to the total flux in the fitted region (354--741 nm). I assigned distinct kinematics to each component: $(V_1, \sigma_1) = (-150, 100)$ $\kms$ for the old population and $(V_2, \sigma_2) = (150, 200)$ $\kms$ for the young one.

I then used the Penalized PiXel Fitting (\ppxf) software\footnote{\label{note:software}Python version 9.4 from \url{https://pypi.org/project/ppxf/}} \citep{Cappellari2004,Cappellari2017,Cappellari2023} to recover the four kinematic parameters $(V_1, \sigma_1, V_2, \sigma_2)$. To explore the parameter space, I coupled \ppxf\ with the Adaptive Metropolis MCMC sampler by \citep{Haario2001} as implemented in the \textsc{AdaMet} package\footnote{Python version 2.0 from \url{https://pypi.org/project/adamet/}} \citep{Cappellari2013p15}. Assuming a uniform prior, the posterior probability is $P\propto\exp(-\chi^2/2)$. I ran a chain of $10^5$ steps for two cases: a low-$\sn$ case ($\sn=2$ per pixel) and a high-$\sn$ case ($\sn=20$).  

The results are shown in \autoref{fig:corner_plot}. At $\sn=2$ (left panel), the posterior distribution is highly complex, non-Gaussian, and biased away from the true input values (magenta lines). The posterior is hard to interpret and one cannot meaningfully quote a single best-fitting value or its error. Averaging such biased results from many low-$\sn$ spaxels would not recover the true average kinematics. In contrast, at $\sn=20$ (right panel), the posterior is unimodal, symmetric, and correctly centred on the true values. The right panel unambiguously reveals the two kinematic components and the parameters are well-constrained. This demonstrates that reaching a sufficient $\sn$ by binning is essential before attempting such a measurement.

\subsection{Star Formation History from Full-Spectrum Fitting} 

My second example concerns the recovery of a galaxy's star formation history (SFH), a problem that involves fitting a spectrum with a large linear combination of template spectra representing stellar populations of different ages and metallicities.
       
Here, I constructed a synthetic galaxy spectrum with a simple SFH, dominated by a single burst of star formation 0.3 Gyr ago. The mock spectrum was built from a linear combination of 25 solar-metallicity templates from the MILES models \citep{Vazdekis2010}, with ages spaced logarithmically between 0.063 and 15.8 Gyr. I then attempted to recover the SFH using \ppxf\ at three different $\sn$ levels: 5, 50, and 500. For each $\sn$ level, I ran 1000 Monte Carlo realisations, adding appropriate Gaussian noise to the spectrum in each run and fitting for the template weights. 
The SFH recovery was performed with a low level of regularization (`regul=10'; see \citealt[fig.~5]{Cappellari2023} for visual reference). This example is not intended to illustrate the expected performance of \ppxf\ in a realistic setting; for that, the reader is referred to \citet{Woo2024}. Rather, its purpose is to demonstrate a generic feature of non-linear models, particularly those used in astrophysical problems where binning is common: at low $\sn$, the model posteriors can be biased and unrepresentative of the true solution.

\autoref{fig:ppxf_sfh} summarizes the Monte Carlo SFH recoveries. At $\sn=5$ the solutions are dominated by noise and a clear systematic bias: the fitter spuriously assigns weight to old populations and fails to recover the true single-burst history. At $\sn=50$ the correct peak is recovered but with substantial scatter, while only at $\sn=500$ does the ensemble reliably reproduce the input SFH with high fidelity. These tests demonstrate that reaching a problem-dependent minimum $\sn$ is essential for trustworthy spectral inference; adaptive binning is therefore a necessary preprocessing step, not merely a cosmetic choice.

\section{Generalizations of Voronoi Diagrams}
\label{sec:voronoi_generalizations}

\begin{figure*}
    \includegraphics[width=\textwidth]{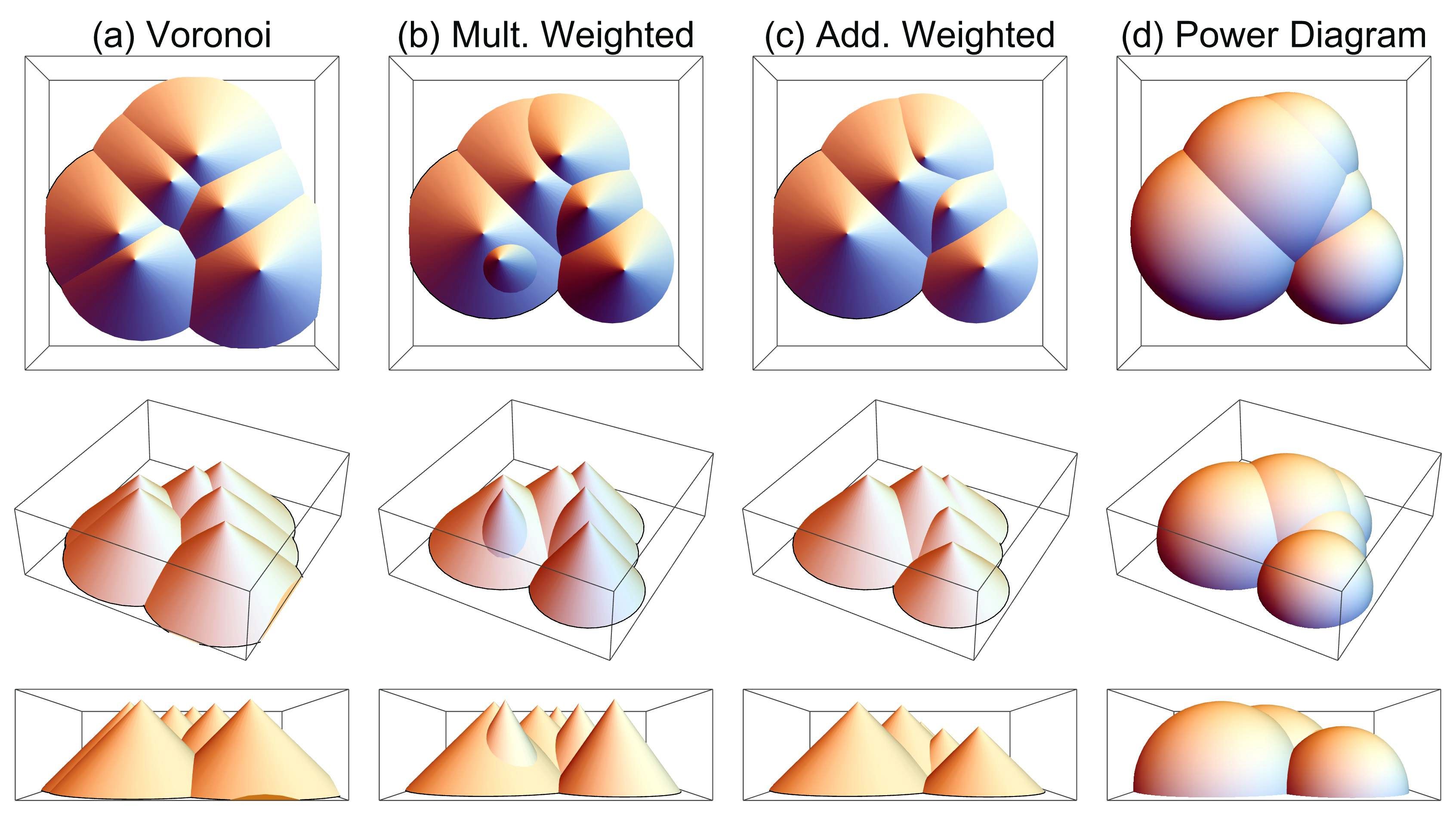} 
\caption{Geometric interpretation of several Voronoi generalizations. Each column illustrates a different tessellation.
The top row shows the view from above: the boundaries of the 2D tessellation are the vertical projections onto the $(x,y)$-plane of the pairwise intersection curves of the 3D surfaces (cones or spheres).
The middle (inclined) and bottom (side-on) rows show the underlying 3D surfaces.
In all cases, the surfaces are centred at the same $(x,y)$ locations (the generators), while their shapes are modified by weights.
(a) \emph{Ordinary Voronoi diagram:} Boundaries come from intersections of identical right circular cones (same slope and same apex height). The projected bisectors are straight lines, yielding convex cells.
(b) \emph{Multiplicatively weighted Voronoi (Apollonius) diagram:} Boundaries come from intersections of cones with different slopes but the same apex height. The weight $w_j$ controls the cone slope, so at any fixed height the base radius scales with $w_j$. The projected bisectors are circular arcs (Apollonius circles), and cells may be non-convex. The figure shows an example where one small cell is contained entirely within a larger one.
(c) \emph{Additively weighted Voronoi (Johnson–Mehl) diagram:} Boundaries come from intersections of cones with the same slope but different apex heights. The weight $w_j$ is encoded in the apex height, and $w_j$ still represents the radius of the cone base. The projected bisectors are hyperbolic arcs, and cells are not guaranteed to be convex. In this example, one cone lies entirely below another, so its corresponding cell is empty and the tessellation has one fewer region than in (a) or (b).
(d) \emph{Power (Laguerre) diagram:} Boundaries come from intersections of spheres centred at $(\mathbf{g}_j,0)$ with radii $r_j=\sqrt{w_j}$. The projected bisectors are straight lines (radical axes), so cells are convex. As in (c), one sphere lies entirely below another, so its corresponding cell is empty and the tessellation has one fewer region.}
\label{fig:weighted_voronoi_3d}
\end{figure*}
 
The original Voronoi-binning algorithm \citep{Cappellari2003} used the ordinary Voronoi tessellation, the simplest member of a broader family of weighted Voronoi diagrams. Later, \citet{Diehl2006} extended the regularization phase by adopting the multiplicatively weighted variant. In order to motivate the adoption of a different tessellation for adaptive binning, it is useful to place these and other variants side-by-side. This section therefore reviews the principal types of weighted Voronoi diagrams, following the classic treatments of \citet[sec.~3]{Okabe2000} and \citet{Aurenhammer2013}, and uses \autoref{fig:weighted_voronoi_3d} to give a geometric interpretation of each. This comparison will make clear that one particular member of the family occupies a special position for our purposes.

\subsection{Ordinary Voronoi Diagram}

Given a set of $n$ distinct points $\mathbf{G}=\{\mathbf{g}_1,\ldots,\mathbf{g}_n\}$ in a $d$-dimensional Euclidean space $\mathbb{R}^d$, called generators, the ordinary Voronoi diagram is a partition of the space into regions based on the nearest-neighbour rule. The Voronoi cell $\mathcal{V}(\mathbf{g}_j)$ associated with a generator $\mathbf{g}_j$ contains all points in $\mathbb{R}^d$ that are closer to $\mathbf{g}_j$ than to any other generator $\mathbf{g}_k$. Using the Euclidean distance 
\begin{equation}
    d(\mathbf{x},\mathbf{g}_j)=\norm{\mathbf{x} - \mathbf{g}_j}
\end{equation}
the cell is defined as:
\begin{equation}
    \mathcal{V}(\mathbf{g}_j)=\{\mathbf{x} \mid \norm{\mathbf{x} - \mathbf{g}_j}\leq \norm{\mathbf{x} - \mathbf{g}_k}, \forall k \neq j\}.
\end{equation}
The boundaries of the Voronoi cells are formed by segments of perpendicular bisectors between pairs of generators. Consequently, the cells are always convex polytopes. Efficient algorithms exist for computing ordinary Voronoi diagrams, with a worst-case time complexity of $\mathcal{O}(n\log n)$ in 2D \citep[e.g., Chapter 3 of][]{Aurenhammer2013}.

Geometrically, the ordinary Voronoi diagram can be visualized as the projection onto the 2D plane of the intersections of a set of identical 3D cones, whose apices are located at the generator positions (\autoref{fig:weighted_voronoi_3d}a). The diagram can be seen by looking at the cones from above.

\subsection{Multiplicatively Weighted Voronoi Diagram}

The multiplicatively weighted Voronoi (MWV) diagram assigns a weight $w_j > 0$ to each generator $\mathbf{g}_j$, with the weight acting as a scaling factor on the distance. This allows one to change the influence of each generator on the partitioning of space. The weighted distance is 
\begin{equation}
d_{\rm MW}(\mathbf{x}, \mathbf{g}_j) = \norm{\mathbf{x} - \mathbf{g}_j} / w_j.
\end{equation}
The cell is defined as:
\begin{equation}
\mathcal{V}_{\rm MW}(\mathbf{g}_j)=\{\mathbf{x} \mid \norm{\mathbf{x} - \mathbf{g}_j}/w_j \leq \norm{\mathbf{x} - \mathbf{g}_k}/w_k, \forall k \neq j\}.
\end{equation}
The boundary between two cells, $\norm{\mathbf{x} - \mathbf{g}_j}/\norm{\mathbf{x} - \mathbf{g}_k} = w_j/w_k$, is a hypersphere (known as a Circle of Apollonius in 2D). A key characteristic of MWV diagrams is that their cells are not necessarily convex and in general can be disconnected. \autoref{fig:weighted_voronoi_3d}(b) shows an example of a MWV diagram with a cell contained inside another. Crucially, the optimal computation time for its generation scales as $\mathcal{O}(n^2)$ \citep[sec.~7.4.2]{Aurenhammer1984,Aurenhammer2013}.

The geometric interpretation of the boundaries of a MWV diagram is the projection of the intersections of cones with different inclinations (slopes), determined by their weights $w_j$ (\autoref{fig:weighted_voronoi_3d}b).

\subsection{Additively Weighted Voronoi Diagram}

The additively weighted Voronoi (AWV) diagram also assigns a real-valued weight $w_j$ to each generator $\mathbf{g}_j$. The distance metric is modified by subtracting this weight, defining a weighted distance 
\begin{equation}
    d_{\rm AW}(\mathbf{x}, \mathbf{g}_j) = \norm{\mathbf{x} - \mathbf{g}_j} - w_j.
\end{equation}
The corresponding cell is:
\begin{equation}
\mathcal{V}_{\rm AW}(\mathbf{g}_j)=\{\mathbf{x} \mid \norm{\mathbf{x} - \mathbf{g}_j} - w_j \leq \norm{\mathbf{x} - \mathbf{g}_k} - w_k, \forall k \neq j\}.
\end{equation}
The boundary between two cells, defined by $\norm{\mathbf{x} - \mathbf{g}_j} - \norm{\mathbf{x} - \mathbf{g}_k} = w_j - w_k$, is a sheet of a hyperboloid of revolution. In 2D, the boundaries are hyperbolic arcs. Like for MWV diagrams, the cells of AWV diagrams are not guaranteed to be convex. The computational complexity is like for the ordinary Voronoi diagrams, scaling as $\mathcal{O}(n \log n)$ \citep[sec.~7.4.1]{Fortune1986,Aurenhammer2013}.

This diagram can be visualized as the projection of the intersections of cones with identical slopes but different heights, where the apex of each cone is shifted vertically by its weight $w_j$ (\autoref{fig:weighted_voronoi_3d}c).

\subsection{Power Diagram: The Ideal Candidate}

Among the family of weighted Voronoi diagrams, the power diagram (also known as the Laguerre-Voronoi diagram) is not just another variation; it possesses a unique combination of geometric and computational properties that make it the ideal mathematical structure for the adaptive binning problem. It is defined using the `power distance', where each generator $\mathbf{g}_j$ is associated with a real-valued weight $w_j$. The power of a point $\mathbf{x}$ with respect to a generator $\mathbf{g}_j$ is given by:
\begin{equation}
\mathrm{pow}(\mathbf{x}, \mathbf{g}_j) = \norm{\mathbf{x} - \mathbf{g}_j}^2 - w_j.
\end{equation}
The power cell $\mathcal{V}_{\rm pow}(\mathbf{g}_j)$ consists of all points whose power with respect to $\mathbf{g}_j$ is less than or equal to their power with respect to any other generator:
\begin{equation}
\mathcal{V}_{\rm pow}(\mathbf{g}_j)=\{\mathbf{x} \mid \norm{\mathbf{x} - \mathbf{g}_j}^2 - w_j \leq \norm{\mathbf{x} - \mathbf{g}_k}^2 - w_k, \forall k \neq j\}.
\end{equation}
The boundary condition, $\norm{\mathbf{x} - \mathbf{g}_j}^2 - \norm{\mathbf{x} - \mathbf{g}_k}^2 = w_j - w_k$, simplifies to a linear equation. This means the boundaries are hyperplanes (straight lines in 2D), which in turn guarantees that the cells are always convex polytopes---a critical property that both AWV and MWV diagrams lack.

Furthermore, this linear boundary property allows power diagrams to be computed with optimal efficiency. By transforming the problem into a convex hull construction in one higher dimension, the tessellation can be found in $\mathcal{O}(n\log n)$ time in 2D \citep{Imai1985,Aurenhammer1987}, matching the speed of the simplest ordinary Voronoi diagram.

Geometrically, the boundaries of a power diagram are the vertical projection of the intersections of a set of upward-opening paraboloids $z = \norm{\mathbf{x} - \mathbf{g}_j}^2 - w_j$. Equivalently, the boundaries of a 2D power diagram can be viewed as the vertical projection of the intersection of a set of 3D spheres with different radii, whose centres lie on the $z=0$ 2D plane (\autoref{fig:weighted_voronoi_3d}d). To see this, consider a sphere $j$ centred at $(\mathbf{g}_j, 0)$ with radius $r_j$. Its equation is $\norm{\mathbf{x} - \mathbf{g}_j}^2 + z^2 = r_j^2$. The intersection of two spheres $j$ and $k$ lies on a plane (the radical plane) defined by equating their equations, which yields $\norm{\mathbf{x} - \mathbf{g}_j}^2 - r_j^2 = \norm{\mathbf{x} - \mathbf{g}_k}^2 - r_k^2$. This is precisely the boundary condition for a power diagram with weights $w_j = r_j^2$. This sphere-based interpretation provides a direct link to the physical analogy of packed soap bubbles (\autoref{fig:honeycomb}), which inspires the fast heuristic algorithm presented in this paper. As I will show in \autoref{sec:cpd_ot}, this specific geometric form makes the power diagram the natural solution to the problem of optimal transport, providing a rigorous mathematical foundation for capacity-constrained binning. This unique combination of guaranteed convexity, computational efficiency, and a direct link to optimal transport theory sets the power diagram apart as the ideal choice for our application.

\section{Centroidal Power Diagrams for Optimal-Transport Binning}
\label{sec:cpd_ot}

In this section, I show that among the various generalizations of Voronoi diagrams, power diagrams are uniquely suited for the adaptive binning of empirical data. This is because they provide a natural solution to a class of problems known as optimal transport, which offers a rigorous mathematical foundation for the binning criteria I outlined in \autoref{sec:intro}.

\subsection{The Optimal Transport Problem}

The theory of optimal transport, first formulated by \citet{Monge1781}, provides a mathematical framework for finding the most efficient way to remap one distribution of `mass' (or any density) to another, given a specified transport cost. For comprehensive reviews of the theory and its computational methods, see \citet{Levy2018} and \citet{Peyre2019}.

For the adaptive binning problem, I consider the `semi-discrete' case: transporting a continuous density distribution $\rho(\mathbf{x})$, which is approximated by our $N$ pixels of data, to a discrete set of $n$ target locations, the bin generators $\{\mathbf{g}_j\}_{j=1}^n$. I define the transport cost as the squared Euclidean distance, $\norm{\mathbf{x} - \mathbf{g}_j}^2$. The goal is to find a partition of the data into a set of $n$ bins $\{\mathcal{V}_j\}$ that minimizes the total transport cost,
\begin{equation}
\mathcal{E}(\{\mathbf{g}_j\}, \{\mathcal{V}_j\}) = \sum_{j=1}^n \int_{\mathcal{V}_j} \norm{\mathbf{x}-\mathbf{g}_j}^2\,\rho(\mathbf{x})\,\dd\mathbf{x},
\label{eq:primal_energy}
\end{equation}
while ensuring each bin $j$ contains a prescribed amount of mass, or `capacity', $\nu_j$. That is, the partition must satisfy the constraint
\begin{equation}
m_j = \int_{\mathcal{V}_j} \rho(\mathbf{x})\,\dd\mathbf{x} = \nu_j, \quad \forall j.
\label{eq:capacity_constraint}
\end{equation}
It can be shown that the optimal partition $\{\mathcal{V}_j\}$ for this problem is a power diagram \citep[e.g.,][]{Aurenhammer1998}.

\subsection{Energy Functional and Centroidal Power Diagrams}

While the primal energy $\mathcal{E}$ in \autoref{eq:primal_energy} is intuitive, finding the partition $\{\mathcal{V}_j\}_{j=1}^n$ that minimizes it under capacity constraints is difficult. The problem becomes tractable by considering the dual problem, which can be expressed using a Lagrangian functional $\mathcal{F}$ that depends on the generator positions $\{\mathbf{g}_j\}$ and a set of real-valued weights $\{w_j\}$ acting as Lagrange multipliers \citep{Aurenhammer1998,Merigot2011,DeGoes2012,Levy2015}:
\begin{equation}
\mathcal{F}(\{\mathbf{g}_j\}, \{w_j\}) = \mathcal{E}(\{\mathbf{g}_j\}, \{\mathcal{V}_j\}) - \sum_{j=1}^n w_j (m_j - \nu_j).
\label{eq:dual_energy}
\end{equation}
Here, $\mathcal{E}$ is the primal energy from \autoref{eq:primal_energy}, where the partition $\{\mathcal{V}_j\}$ is now explicitly shown to be the power diagram defined by the weights $\{w_j\}$. The term $m_j$ is the current capacity of cell $\mathcal{V}_j$ from \autoref{eq:capacity_constraint}, and $\nu_j$ is the target capacity, which for the Voronoi-binning problem I generally assume to be constant $\nu$, although this is not a requirement for the method.

The key insight is that finding the optimal binning is equivalent to finding a saddle point of $\mathcal{F}$. The gradients of $\mathcal{F}$ with respect to the weights and generator positions reveal its utility:
\begin{enumerate}
    \item \textbf{Gradient w.r.t. weights:} The gradient with respect to a weight $w_j$ is simply the difference between the cell's target capacity and its current capacity \citep{Aurenhammer1998,DeGoes2012}:
    \begin{equation}
        \nabla_{w_j} \mathcal{F} = \nu_j - m_j.
    \end{equation}
    For a fixed set of generators, finding the weights $\{w_j\}$ that maximize the dual functional $\mathcal{F}$ is a convex optimization problem \citep{Aurenhammer1998}. This is a crucial property, as it guarantees the existence of a unique global maximum. Standard and efficient algorithms, such as Newton's method, can be used to find this solution by driving the gradient to zero, thus ensuring the capacity constraints $m_j = \nu_j$ are satisfied.

    \item \textbf{Gradient w.r.t. generators:} The gradient with respect to a generator position $\mathbf{g}_j$ is \citep{DeGoes2012}:
    \begin{equation}
        \nabla_{\mathbf{g}_j} \mathcal{F} = 2 m_j (\mathbf{g}_j - \mathbf{b}_j),
    \end{equation}
    where $\mathbf{b}_j$ is the barycentre (density-weighted centroid) of the cell $\mathcal{V}_j$. Setting this gradient to zero implies that the generator must coincide with its cell's barycentre: $\mathbf{g}_j = \mathbf{b}_j$.
\end{enumerate}
A configuration that is a stationary point of $\mathcal{F}$---simultaneously satisfying the capacity constraints and the barycentric condition---is called a \emph{Centroidal Power Diagram} (CPD). A CPD corresponds to a (local) minimum of the original transport energy $\mathcal{E}$, thus providing a complete and principled solution to the adaptive binning problem.

\subsection{Challenges in Applying to Astronomical Data}

While the CPD framework is theoretically ideal, its direct application to the binning of astronomical data faces two main practical challenges:
\begin{enumerate}
  \item \textbf{Discrete Data vs. Continuous Theory:} The optimal transport theory is formulated for continuous density functions. When applied to discrete data, integrals are replaced by sums over pixels. This approximation is valid when bins are large, but it breaks down when bins contain only a few pixels, leading to numerical instabilities. While one could interpolate the discrete data to create a continuous density, this is only rigorously applicable when the capacity function is additive.

  \item \textbf{Non-Additive Capacity:} The most significant challenge is that the bin `capacity' is often not a simple additive quantity. For example, when binning to a target signal-to-noise ratio ($\sn$), the bin's total $(\sn)^2$ is only the sum of the pixel $(\sn)^2$ if the noise is uncorrelated \citep[see][sec.~2]{Cappellari2003}. In practice, instrumental effects and data reduction steps (e.g., dithering, resampling) introduce significant covariance between pixels \citep[see][sec.~6.2]{Westfall2019}. This makes the capacity $m_j$ a non-linear, non-additive function of its constituent pixels. As a consequence, the dual functional $\mathcal{F}$ loses its convenient convexity at fixed generator positions, and the analytic gradients become invalid. As a result, standard gradient-based optimization methods become unstable and fail to converge.
\end{enumerate}

I confirmed this limitation through numerical experiments using the formalisms of \citet{Aurenhammer1998} and \citet{DeGoes2012}. While these variational approaches perform well for additive capacities in the continuum limit (i.e., large bins with many pixels), they fail catastrophically for the non-additive capacities typical of real data with correlated noise.

Because of these issues, a direct implementation of a mathematically exact CPD solver is not robust for real-world data binning. In the following section, I introduce a new algorithm, \textsc{PowerBin}, which is inspired by the optimal transport framework but uses a fast and robust heuristic to handle these complexities.

\section{Fast centroidal power-diagram solver}
\label{sec:fast_power_insight}

\begin{figure*}
        \includegraphics[width=0.49\textwidth]{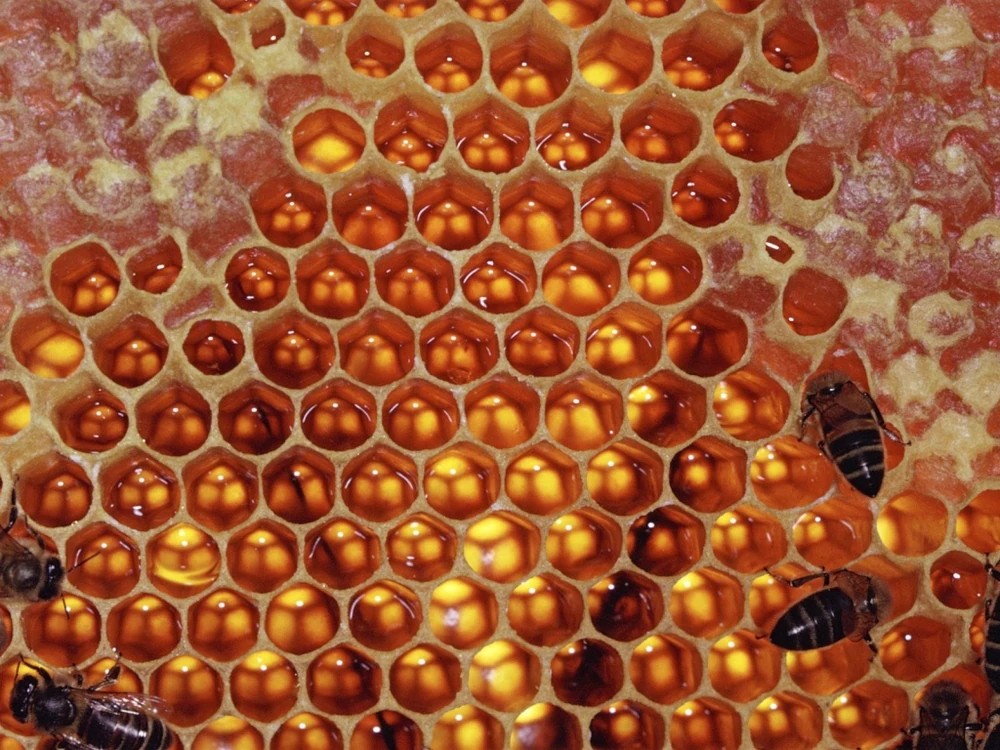}
        \includegraphics[width=0.49\textwidth]{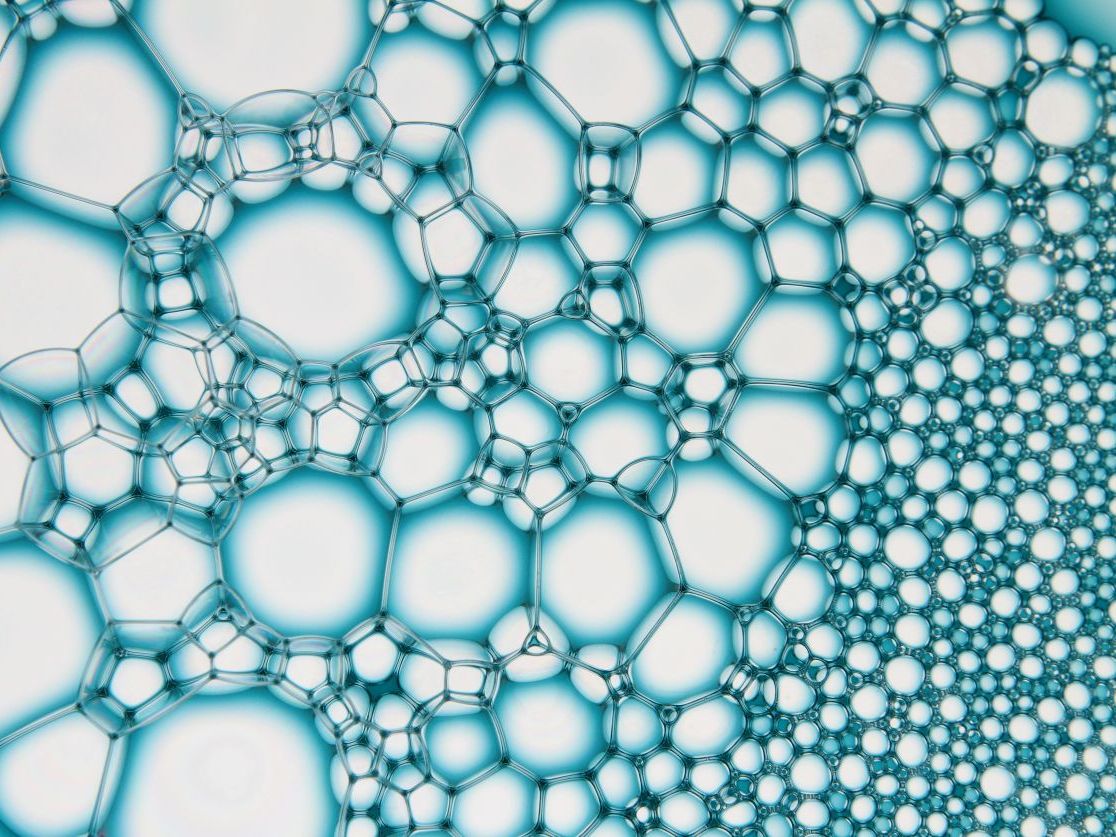}
        \caption{Natural analogues for optimal data binning, illustrating the principles of compactness and capacity uniformity. \textit{Left:} A \href{https://beekeeping.fandom.com/wiki/Honey}{honeycomb} demonstrates the optimal packing of equal-area cells (hexagons). \textit{Right:} A 2D foam (soap bubbles) provides a physical model for a capacity-constrained tessellation. By minimizing surface energy, the bubbles form a structure equivalent to a power diagram, where different bubble sizes correspond to different cell capacities. This illustrates the physical principle behind the \textsc{PowerBin} algorithm. Image courtesy of \href{https://lincolnmathsphys.wordpress.com/2015/02/26/fascinating-foams-the-mathematics-of-soap-bubbles/}{Professor Simon Cox}, Aberystwyth University.
        \label{fig:honeycomb}}
\end{figure*}

The previous section established that while Centroidal Power Diagrams (CPDs) provide a theoretically ideal framework for adaptive binning, formal solvers based on gradient-based optimization of the dual energy functional are impractical for real astronomical data. The non-additive nature of capacity measures like $\sn$ with correlated noise violates the assumptions required for these methods to converge reliably. This section introduces the core of the \textsc{PowerBin} algorithm: a fast, robust, and physically-motivated heuristic that bypasses these problems. Instead of relying on complex and fragile numerical optimization, I develop a simple iterative scheme inspired by the geometry of packed cells, which proves highly effective at enforcing capacity constraints while maintaining computational efficiency and bin convexity.

\subsection{A physical heuristic for the weight–capacity relation}

The central challenge in constructing a capacity-constrained power diagram is to find the set of weights $\{w_j\}$ that yields cells with the desired capacities $\{\nu_j\}$. As noted by experts in the field, `the relation between the weights and the measures of the power cells is non-trivial' \citep[Sec.~2.4]{Levy2015}. This complexity arises from two main factors. First, the weights are defined only up to a common additive constant; adding the same value to all weights leaves the tessellation unchanged. Second, for an arbitrary arrangement of generators, there is no simple, direct relationship between a weight $w_j$ and the area $A_j$ of its corresponding cell. The cells can be highly elongated, and some generators may even have empty cells.

However, the problem simplifies dramatically if one considers the specific geometry of a \emph{centroidal} tessellation, where each generator is close to the centre of its cell. My approach is based on two key insights:
\begin{enumerate}
    \item \textbf{Power weights as squared radii.} As shown in \autoref{sec:voronoi_generalizations}, a power diagram can be defined by associating a circle of radius $r_j$ with each generator $\mathbf{g}_j$, such that the weight is $w_j = r_j^2$. This provides a direct geometric interpretation: the cell boundaries are the radical axes of these circles, and their sizes control the tessellation.

    \item \textbf{The `packed bubbles' approximation.} In a centroidal configuration, the generators are close to the cell centroids, which naturally produces compact, nearly-round cells. In this limit, the tessellation resembles a foam of packed bubbles (\autoref{fig:honeycomb}). For such a packing, the area $A_j$ of a cell is well-approximated by the area of its defining circle:
    \begin{equation}
        A_j \approx \pi r_j^2.
        \label{eq:area_approx}
    \end{equation}
    This simple approximation provides the crucial link between a cell's area ($A_j$) and its generator's weight ($w_j = r_j^2$).
\end{enumerate}

With this physical model, I can derive a simple update rule. The goal is to adjust the radii $\{r_j\}$ until the measured capacity $m_j$ of each cell matches a target value $\nu$. I start by assuming that a cell's capacity is, to first order, proportional to its area: $m_j \approx \rho_j A_j$, where $\rho_j$ is an effective local capacity density.

To achieve the target capacity $\nu$, cell $j$ would need a target area $A_j^{\star} \approx \nu / \rho_j$. Using our approximation from \autoref{eq:area_approx}, the corresponding target squared radius would be $r_j^{2\,\star} \approx A_j^{\star}/\pi = \nu/(\pi\,\rho_j)$. The unknown density $\rho_j$ can be eliminated by substituting its value from the current iteration, $\rho_j \approx m_j/A_j$. This yields a simple update rule for the target squared radius based entirely on measurable quantities from the current tessellation:
\begin{equation}
r_j^{\text{new}} \leftarrow \sqrt{\frac{\nu}{m_j} \frac{A_j}{\pi}} = \sqrt{\frac{f_j A_j}{\pi}}, \quad \text{where} \quad f_j \equiv \frac{\nu}{m_j}.
\label{eq:multiplicative_rule}
\end{equation}

In terms of the power weights themselves, the update is $w_j^{\text{new}} \leftarrow f_j A_j / \pi$.

This leads to the simple yet powerful iterative algorithm summarized in \autoref{alg:powerbin_code}. The update rule in \autoref{eq:multiplicative_rule} is formally similar to the WVT rule of \citet{Diehl2006}, but it operates on the radii of a power diagram, not the weights of an MWV diagram. This is a crucial distinction: for MWV diagrams, multiplying all weights by a common factor has no effect, whereas for a power diagram, it changes the tessellation. This implies that for a MWV the constant factors are irrelevant, while for a Power Diagram they are essential. By combining the natural area–radius relation ($A\approx \pi r^2$) with centroidal recentring, this heuristic achieves stable, few-iteration convergence for capacity equalization while preserving the convexity and compactness of the bins.

\begin{algorithm}[t]
\caption{PowerBin}
\label{alg:powerbin_code}
\begin{algorithmic}
\Require spaxels $\{\mathbf{x}_i\}_{i=1}^N$, capacity function $\mathcal{C}(I)$, target $\nu$
\Ensure Bin map $\{b_i\}$, generators $\{\mathbf{g}_j\}$, radii $\{r_j\}$
\Statex
\State \textbf{Initialization:}
\State Estimate \texttt{pixelsize} if not provided
\State $\{\mathbf{x}_i\} \gets \{\mathbf{x}_i\} / \texttt{pixelsize}$ \Comment{Normalize coordinates}
\State $\{\mathbf{g}_j\} \gets \textsc{BinAccretion}(\{\mathbf{x}_i\}, \nu, \mathcal{C})$ \Comment{\autoref{alg:powerbin_accretion}}
\State $r_j \gets 1$ \Comment{Initialize power-cell radii for all bins $j$}
\Statex
\State \textbf{Bins Regularization:}
\For{$t=1$ to \texttt{maxiter}}
  \State $\mathbf{g}^{\rm old}_j \gets \mathbf{g}_j$; \quad $r^{\rm old}_j \gets r_j$ \Comment{Store state for all bins $j$}
  \State $(\{\mathbf{g}_j\},\{A_j\},\{m_j\}) \gets\textsc{UpdateBins}(\{\mathbf{x}_i\},\{\mathbf{g}_j\},\{r_j\},\mathcal{C})$
  \State $r_j \gets \sqrt{\nu A_j/(m_j \pi)}$ \Comment{For each bin $j$}
  \If{oscillation detected}
    \State $\mathbf{g}_j \gets \mathbf{g}^{\rm old}_j + d \cdot (\mathbf{g}_j - \mathbf{g}^{\rm old}_j)$ \Comment{Apply to all $j$}
    \State $r_j \gets r^{\rm old}_j + d \cdot (r_j - r^{\rm old}_j)$ \Comment{Under-relaxation $d \in (0,1)$}
  \EndIf
  \State $d_{j,k} \gets \norm{\mathbf{g}_j - \mathbf{g}_k}$ for nearest neighbour $k$ \Comment{For each bin $j$}
  \State $r_j \gets \max(0.5, \min(r_j, d_{j,k} - 0.5))$ \Comment{Prevent empty cells}
  \State $\Delta\gets \mathrm{norm}(\{\mathbf{g}_j-\mathbf{g}^{\rm old}_j\})$
  \If{$\Delta$ is small or cycling} \State \textbf{break} \EndIf
\EndFor
\Statex
\State \textbf{Finalize:}
\State $\{b_i\} \gets \textsc{PowerDiagram}(\{\mathbf{x}_i\},\{\mathbf{g}_j\},\{r_j\})$
\State Rescale $\{\mathbf{g}_j\}$ and $\{r_j\}$ by \texttt{pixelsize}
\State \Return $\{b_i\}, \{\mathbf{g}_j\}, \{r_j\}$
\end{algorithmic}
\end{algorithm}

\begin{algorithm}[t]
\caption{UpdateBins}
\label{alg:powerbin_updatebins}
\begin{algorithmic}
\Require Spaxels $\{\mathbf{x}_i\}$, generators $\{\mathbf{g}_j\}$, radii $\{r_j\}$, capacity $\mathcal{C}(I)$
\Ensure Updated centroids $\{\mathbf{g}_j\}$, areas $\{A_j\}$, capacities $\{m_j\}$

\State $\{b_i\} \gets \textsc{PowerDiagram}(\{\mathbf{x}_i\},\{\mathbf{g}_j\},\{r_j\})$
\State $I_j \gets \{i \mid b_i=j\}$; \quad $A_j \gets |I_j|$ \Comment{For each bin $j$}
\State $\mathbf{g}_j \gets \mathrm{mean}\{\mathbf{x}_i \mid i\in I_j\}$
\State $m_j \gets C(I_j)$
\State \Return $\{\mathbf{g}_j\},\{A_j\},\{m_j\}$
\end{algorithmic}
\end{algorithm}

\begin{algorithm}[t]
\caption{PowerDiagram}
\label{alg:powerbin_powertess}
\begin{algorithmic}
\Require Spaxels $\{\mathbf{x}_i\}$, generators $\{\mathbf{g}_j\}$, radii $\{r_j\}$
\Ensure Bin map $\{b_i\}$

\State $r_{\max} \gets  (1+\varepsilon)\cdot\max_j\{r_j\}$
\State $z_j \gets \sqrt{r_{\max}^2-r_j^2}$ \Comment{For each generator $j$}
\State build KDTree on lifted points $\{(\mathbf{g}_j,z_j)\} \in \mathbb{R}^3$
\State $b_i \gets$ NN of $(\mathbf{x}_i,0)$ in KDTree \Comment{For each pixel $i$}
\State \Return $\{b_i\}$
\end{algorithmic}
\end{algorithm} 

\subsection{Implementation Details}
\label{sec:implementation_details}

The success and speed of the regularization stage of the \textsc{PowerBin} algorithm rely on a few crucial implementation choices, which are detailed in \autoref{alg:powerbin_code}, \ref{alg:powerbin_updatebins} and \ref{alg:powerbin_powertess}.

\paragraph*{Efficient Power Tessellation:}
A key advantage of power diagrams over other weighted Voronoi diagrams is their computational efficiency. A power diagram can be computed with $\mathcal{O}(N\log N)$ complexity \citep{Aurenhammer1987}, whereas an MWV diagram requires $\mathcal{O}(N^2)$ operations \citep{Aurenhammer1984}. I achieve this efficiency for my discrete dataset by implementing the geometric lifting technique described in \citet[Sec.~5]{Imai1985}, as detailed in \autoref{alg:powerbin_powertess}. Each 2D generator $\mathbf{g}_j$ with radius $r_j$ is `lifted' to a 3D point $(\mathbf{g}_j, z_j)$ where $z_j^2 = r_{\max}^2 - r_j^2$. The power-diagram assignment for a 2D spaxel $\mathbf{x}_i$ is then equivalent to finding the nearest 3D neighbour to the point $(\mathbf{x}_i, 0)$ among the lifted generators. This 3D nearest-neighbour search is performed efficiently using a standard KD-Tree (\href{https://docs.scipy.org/doc/scipy/reference/generated/scipy.spatial.KDTree.html}{scipy.spatial.KDTree}) in the SciPy library \citep{Scipy2020}, which implements the algorithm of \citet{Maneewongvatana1999}.

\paragraph*{Geometric Centroids vs. Barycentres:}
A key choice in my algorithm is the update of the generators. Instead of moving them to the capacity-weighted barycentre of each cell, I update them to the cell's unweighted, geometric centroid (the mean position of its constituent pixels). This means I do not strictly minimize the optimal transport energy functional from \autoref{eq:primal_energy}. However, this choice is deliberate and crucial for robustness. The main reason is that it allows the method to handle data with negative values (e.g., background-subtracted X-ray data), where the definition of a barycentre breaks down. This choice was adopted in the standard \textsc{VorBin} algorithm for the same reason. Moreover, this choice makes the algorithm a more direct implementation of the physical `soap bubble' analogy (\autoref{fig:honeycomb}) that inspired my area-weight relation, as the cells are centred geometrically rather than by mass. Furthermore, for real data with non-additive capacities, the energy functional itself is ill-defined, making its formal minimization moot. For positive data, I find that my algorithm is not particularly sensitive to this choice, unlike the standard unweighted CVT which relies on the barycentric condition.

\paragraph*{Failure of Formal Optimization:}
The use of a heuristic update rule and geometric centroids is a direct consequence of the practical failure of formal optimization methods. During development, I implemented a version of the algorithm that directly minimized the energy functional from \citet{Aurenhammer1998} and \citet{DeGoes2012} using their analytic gradient and a quasi-Newton optimizer. This approach is not just slower; it fails completely. Even in the continuum limit with many spaxels per bin, the method fails to converge to a sensible result (i.e., the capacity is not equalized) when the capacity function is non-additive, as is generally the case with real data. The fast, physically-motivated heuristic of \textsc{PowerBin} proved to be far superior in all practical tests, converging quickly and reliably to a high-quality solution where formal methods could not.

\paragraph*{Robust Convergence:}
Iterative methods can sometimes stall or enter a cycle. To prevent this, I use under-relaxation when I detect cycling, and employ a robust early-stopping heuristic. This method monitors the sequence of generator shifts and terminates the loop if the improvement stagnates or if it detects that the values are oscillating without any significant downward trend. This ensures reliable termination even in difficult cases.

\paragraph*{Generality of Capacity Function:}
The combination of my robust heuristic and the use of geometric centroids makes the \textsc{PowerBin} algorithm extremely versatile. The capacity function is not limited to the specific forms discussed in this paper and can be adapted to various applications. This flexibility allows for the incorporation of complex, non-additive constraints, making the algorithm applicable to a broad range of problems.

\section{A linear-time bin-accretion algorithm}
\label{sec:accretion}

\begin{algorithm}[t]
\caption{BinAccretion}
\label{alg:powerbin_accretion}
\begin{algorithmic}
\Require spaxels $\{\mathbf{x}_i\}_{i=1}^N$, capacity function $\mathcal{C}(I)$, target $\nu$
\Ensure Initial bin generators $\{\mathbf{g}_j\}$
\Statex
\State \textbf{Initialization:}
\State $\rho_i \gets \mathcal{C}(\{i\})$ \Comment{For each pixel $i$}
\State $b_i \gets 0$ \Comment{Initialize all pixels $i$ as unbinned}
\State Pre-compute Delaunay adjacency graph for all $\{\mathbf{x}_i\}$
\State Build max-heap of all pixels, prioritized by $\rho_i$
\State $j \gets 1$ \Comment{Initialize bin counter}
\Statex
\State \textbf{Accretion Loop:}
\While{unbinned pixels exist}
    \State Pop unbinned pixel $k$ with max $\rho_k$ from heap as new seed
    \State $I_j \gets \{k\}$; \quad $b_k \gets j$
    \State $m_j \gets \rho_k$; \quad $\mathbf{c}_j \gets \mathbf{x}_k$
    \State $\mathcal{R}_j \gets 0$ \Comment{Initialize roundness metric}
    \State $F \gets$ unbinned Delaunay neighbours of $k$
    \While{$m_j < \nu$ and $F$ is not empty}
        \State Find pixel $p \in F$ closest to centroid $\mathbf{c}_j$
        \State Update $\mathbf{c}_j$ and $\mathcal{R}_j$ with pixel $p$ using Welford's algorithm
        \If{$\mathcal{R}_j$ exceeds roundness threshold} \State \textbf{break} \EndIf
        \State $m_j^{\rm old} \gets m_j$; \quad $m_j \gets \mathcal{C}(I_j \cup \{p\})$
        \If{$m_j + m_j^{\rm old} > 2\nu$} \Comment{Capacity diverging from target} \State \textbf{break} \EndIf
        \State $I_j \gets I_j \cup \{p\}$; \quad $b_p \gets j$
        \State Update frontier $F$ with neighbours of $p$
    \EndWhile
    \State $j \gets j + 1$
\EndWhile
\Statex
\State \textbf{Finalize:}
\State Identify bins that failed to reach a fraction of $\nu$
\State Reassign their pixels to the nearest successful bin
\State Compute final generators $\{\mathbf{g}_j\}$ as centroids of the final bins
\State \Return $\{\mathbf{g}_j\}$
\end{algorithmic}
\end{algorithm}

A crucial, and perhaps under-appreciated, aspect of all successful adaptive-binning schemes is the quality of the initial tessellation. The iterative refinement stages, whether based on a Centroidal Voronoi Tessellation (CVT), a Weighted Voronoi Tessellation (WVT), or the Centroidal Power Diagram (CPD) presented here, are all local optimizers. They are variants of \citet{Lloyd1982} algorithm, which is known to be sensitive to the initial placement of the generators. Unlike the optimization of weights at fixed generator's location, the optimization of the energy functional of \autoref{eq:dual_energy} with respect to the generator positions is not convex and presents a large number of secondary minima \citep[see][fig.~4]{Levy2015}. This means that the final result can vary significantly depending on the starting configuration.

One cannot, for instance, initialize the generators with points drawn randomly from the underlying signal or $\sn$ distribution and expect the iterations to converge to a satisfactory result. The discrete nature of the data and the non-convexity of the optimization landscape mean that such an approach will invariably become trapped in a poor local minimum, yielding a tessellation with suboptimal bin shapes and poor capacity uniformity. Consequently, the bin-accretion algorithm, first introduced in \citet[sec.~4.2]{Cappellari2003}, has always been the indispensable foundation of the entire procedure. It provides an excellent initial guess that already satisfies the capacity constraint, allowing the subsequent refinement to focus solely on improving the bin morphology.

With the development of the fast CPD solver, which has a time complexity of $\mathcal{O}(N\log N)$ for $N$ pixels, the original bin-accretion algorithm became the computational bottleneck. To fully realize the performance gains of the new method, it was essential to devise an accretion algorithm with a comparable, near-linear time complexity. I achieve this through four key improvements, while the complete revised method is described in \autoref{alg:powerbin_accretion}:

\begin{enumerate}
    \item \textbf{Delaunay Adjacency:} I begin by pre-computing a single Delaunay triangulation of all input pixel coordinates. This is an $\mathcal{O}(N\log N)$ operation that provides a static adjacency graph for the entire dataset. For any given pixel, its neighbours are instantly known without requiring any further geometric searches. The computation is done using \href{https://docs.scipy.org/doc/scipy/reference/generated/scipy.spatial.Delaunay.html}{scipy.spatial.Delaunay}, which is based on the QHull library \citep{Barber1996}.

    \item \textbf{Frontier-Based Growth:} To efficiently find the next pixel to add, the algorithm restricts its search to the bin's `frontier'. This frontier is defined as the set of unbinned pixels that are Delaunay neighbours to any pixel already in the current bin. By only considering this small subset of pixels at each step, the search for the one closest to the bin's centroid becomes extremely fast.

    \item \textbf{Incremental Updates:} All quantities required to assess the validity of adding a new pixel to a bin---namely its centroid and mean squared radius---are updated incrementally using \citet{Welford1962} algorithm. The total capacity is also updated incrementally if it is additive; for an additive capacity, this simply means adding the capacity of the new pixel to the running total for the bin. A recomputation for the entire bin is still required in the case of a black-box generic capacity function. Adding a pixel involves a simple update to a running sum, an $\mathcal{O}(1)$ operation, rather than a full re-computation over all pixels in the growing bin.

    \item \textbf{Heap-Managed Seeding:} To efficiently select the starting point for each new bin, I use a max-heap data structure containing all unbinned pixels, prioritized by their initial capacity density (or `brightness', or $\sn$). After a bin is completed, the next seed is retrieved from the top of the heap in an $\mathcal{O}(\log N)$ operation. This approach is more efficient than the original method of \citet{Cappellari2003}, which started a new bin from the unbinned pixel closest to the centroid of all remaining unbinned pixels. The brightness of a pixel is a static property, allowing for a single heap construction at the start, whereas the centroid of unbinned pixels changes after every bin is created, requiring a costly recalculation.
\end{enumerate} 

In many common astronomical applications, such as binning the surface brightness of a galaxy, this change in seeding strategy has a minimal impact on the final result. In such cases, the brightest unbinned pixels are naturally located near the galaxy's centre, which is also where the centroid of the remaining light is. The old method would accrete bins in nearly spherical shells around the galaxy's core, while the new `brightest-pixel' approach will still start bins near the centre but will tend to grow them along isophotes. In general, both methods provide an excellent starting point for the subsequent regularization. However, the new approach appears superior in the case of multiple distinct objects, as illustrated in \autoref{sec:background_limited}. 

\begin{figure*}
  \centering
  \includegraphics[width=0.83\textwidth]{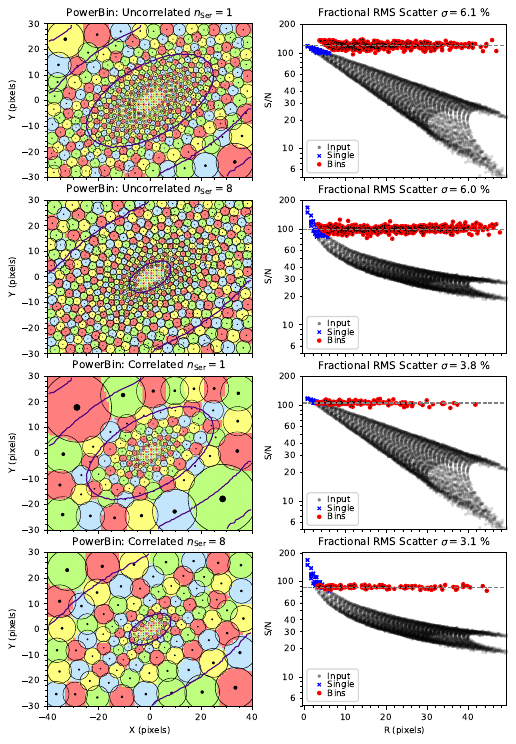}
\caption{The performance of the \textsc{PowerBin} algorithm is shown for two mock galaxies ($n_{\rm Ser}=1$ and $n_{\rm Ser}=8$). Both simulations have $R_{\rm eff}=10$ pixels and an axial ratio of $q'=0.5$. The target signal-to-noise ratio ($\snt$) was calibrated to ensure $\approx 40$ unbinned spaxels near the centre. \textit{Left:} Resulting Power Diagrams; circle radii are $r_j = \sqrt{w_j}$, with small black circles marking the generator points. Contours show the galaxy's $\sn$, spaced logarithmically by 1 mag. \textit{Right:} $\sn$ distribution. Original spaxel $\sn$ values are grey circles; $\snt$ is the dashed line. Blue crosses are single spaxels ($ \sn > \snt$), and large red circles are the final bin $\sn$. Top panels use a simple Poissonian noise model; bottom panels show robustness to non-additive capacities from correlated noise. In all cases, the algorithm optimizes the capacity function $\mathcal{C}=(\sn)^2$ (which is an additive capacity in the Poissonian case), but I plot the square root $\sqrt{\mathcal{C}}=\sn$.} \label{fig:powerbin_sersic_stack}
\end{figure*}

\begin{figure*}
  \centering
  \includegraphics[width=\textwidth]{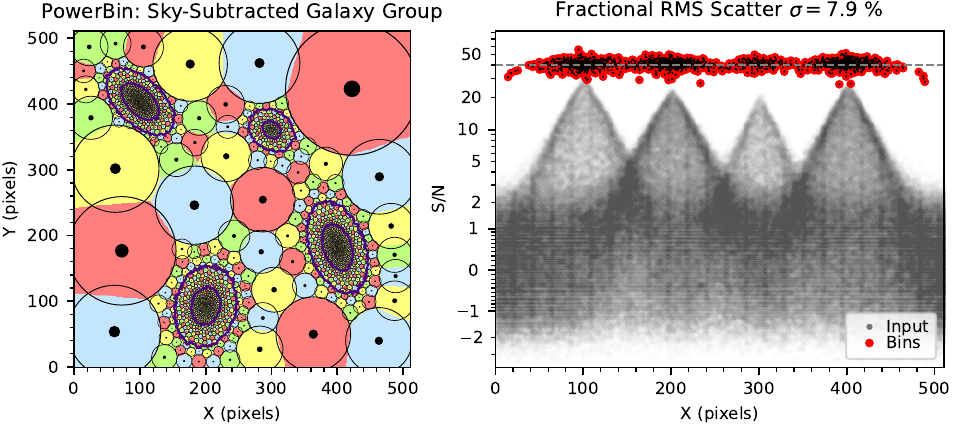}
  \caption{Application of the \textsc{PowerBin} algorithm to a mock galaxy group, simulating a background-limited observation with multiple sources in a large, noisy field. For the large majority of the image spaxels, the signal is essentially zero, with only some nearly constant random noise remaining after imperfect sky subtraction. The plot format is the same as in \autoref{fig:powerbin_sersic_stack}. \textit{Left panel:} The final power diagram tessellation. Circles indicate the bin radii ($r_j=\sqrt{w_j}$), and small black disks mark the generators. Galaxy isophotes are spaced by 1 mag in $\sn$. The algorithm correctly bins the galaxy signal while filling the empty background with large, low-signal bins. \textit{Right panel:} The $\sn$ distribution, showing the original spaxel $\sn$ (gray filled circles), the target $\snt$ (dashed line), and the final bins $\sn$ (red filled circles). The bins are clustered tightly around the target value, with the exceptions of some pure-noise outer bins falling slightly below $\snt$, demonstrating the algorithm's robustness. Here the spaxels capacity can be negative and I equalize the capacity function $\mathcal{C}=\sn$ directly (not squared).}
  \label{fig:powerbin_galaxy_group}
\end{figure*}

Apart from these significant algorithmic optimizations, the new implementation aims to reproduce the logic of the original bin-accretion algorithm from \citet[sec.~5.1]{Cappellari2003}. There are, however, two minor differences in the acceptance criteria. (a) I employ a different definition of roundness, based on the mean squared radius of the pixel coordinates, which is faster to update incrementally. (b) The precise conditions for accepting a new pixel into a bin have been slightly adjusted. The new algorithm therefore represents a much faster, but functionally very similar, version of the well-tested accretion method. It is worth noting that the roundness criterion is not critical for the success of the overall algorithm; the final tessellation after the regularization stage is quite similar, though not identical, even if this test is omitted entirely from the accretion phase.

\section{Power Binning Examples}
\label{sec:applications}

In this section, I demonstrate the compactness and uniformity performance, and the versatility of the \textsc{PowerBin} algorithm. I apply it to a range of test cases, including simulated galaxies with different morphologies and noise properties, real integral-field spectroscopic data, and a large, complex image to showcase its scalability and applicability beyond standard astronomical use cases.

\begin{figure*}
  \centering
  \includegraphics[width=\textwidth]{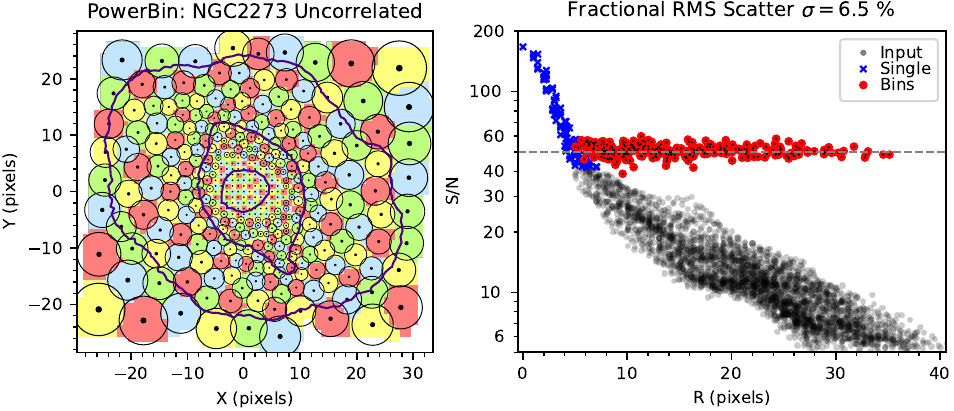}

  \caption{Application of the \textsc{PowerBin} algorithm to the SAURON integral-field data of the galaxy NGC 2273. This dataset was used as the primary test case in the original Voronoi-binning paper \citep{Cappellari2003} and has been included for reference in the public \textsc{VorBin} software package for two decades. The plot format is the same as in \autoref{fig:powerbin_sersic_stack}, assuming uncorrelated noise. \textit{Left panel:} The final power diagram tessellation, with circles indicating the bin radii ($r_j=\sqrt{w_j}$) and black disks marking the generators. Galaxy isophotes are spaced by 1 mag in $\sn$. \textit{Right panel:} The $\sn$ distribution, showing the original spaxel $\sn$ (gray filled circles), the target $\snt$ (dashed line), the unbinned spaxels (blue crosses), and the final bin $\sn$ (red filled circles).}
  \label{fig:powerbin_ngc2273}
\end{figure*}

\subsection{Application to Mock Single Galaxies}

I first test the algorithm on mock galaxy data to assess its quality performance under controlled conditions. I created two types of galaxy images: one with an exponential disk profile (Sérsic $n_{\rm Ser}=1$) and another with a highly concentrated, elliptical-like profile \citep[e.g.,][]{Kormendy2009} described by a \citet{Sersic1968} profile ($n_{\rm Ser}=8$). These two cases test the algorithm's ability to handle both shallow and steep signal gradients. For each galaxy, I consider two scenarios for the noise: uncorrelated or correlated.

The results are shown in \autoref{fig:powerbin_sersic_stack}. To aid visualization, the tessellation is coloured using the \href{https://networkx.org/documentation/stable/reference/algorithms/coloring.html}{networkx.coloring.greedy\_color} algorithm from the NetworkX package \citep{Hagberg2008} to approximately four-colour the Delaunay graph of the bin generators. The top two panels show the ideal case of uncorrelated, Poissonian noise. In this scenario, the bin signal-to-noise is calculated using the standard capacity function for uncorrelated noise
\begin{equation}\label{eq:capacity_uncorrelated}
    \mathcal{C}(\{i\})=(\snb)^2 = \frac{(\sum_i \mathcal{S}_i)^2}{\sum_i \mathcal{N}_i^2}, 
\end{equation}
where $\mathcal{S}_i$ and $\mathcal{N}_i$ are the signal and noise of the individual pixels in the bin. For Poissonian noise, $\mathcal{N}_i^2 = \mathcal{S}_i$, and the bin capacity becomes $(\snb)^2 = \sum_i \mathcal{S}_i$, which is additive. The left sub-panels show the resulting power diagram tessellation. The bins are compact and convex, and their sizes adapt smoothly to the underlying signal gradient, becoming larger in the low-$\sn$ outskirts. The right sub-panels confirm that the final bin $\sn$ (blue points) clusters tightly around the target value (dashed line), demonstrating excellent uniformity.

The bottom two panels illustrate a more realistic scenario where noise covariance is present. I simulate this by making the capacity function non-additive, using an empirical formula derived from real IFS data to penalize the $\sn$ of bins with many pixels. Specifically, the bin $\sn$ is modified as
\begin{equation}
    \snb \rightarrow \frac{\snb}{1 + 1.07 \lg N_{\rm pix}},
\end{equation}  
where $N_{\rm pix}$ is the number of spaxels in the bin. This formula is not general but depends on the data under study. It was derived for CALIFA data \citep[fig.~11]{GarciaBenito2015} and is used here for illustrative purposes. This test highlights a key strength of my heuristic approach: despite the non-linear and non-additive nature of the capacity, the algorithm converges robustly and still produces bins with excellent $\sn$ uniformity. A formal gradient-based optimizer would fail to converge in this regime, but the physically-motivated update rule of \textsc{PowerBin}, combined with the bin-accretion initialization, handles it with ease.

\subsection{Application to Mock Sky-Subtracted Data}
\label{sec:background_limited}

A crucial consideration for modern IFS data is that they are often background-limited rather than purely Poissonian. In this regime, the noise is roughly constant across the field, meaning that accreting spaxels with very low signal can decrease a bin's $\sn$ instead of increasing it. This raises the question of how the algorithm performs on large mosaics that may contain multiple sources of interest within a vast field of mostly `empty sky'.

To address this scenario, I simulated a mock galaxy group image, shown in \autoref{fig:powerbin_galaxy_group}. The simulation generates an image by adding four exponential profile galaxies onto a constant sky background of 100 counts. The galaxy models have peak surface brightness ranging from 500 to 1000 counts. The combined image is then subjected to Poisson noise, and a slightly imperfect sky subtraction is performed to mimic a realistic data reduction process. This creates a background-limited image where the noise is nearly constant, dominated by the sky contribution. The results demonstrate that \textsc{PowerBin} handles this challenging case robustly. It correctly identifies and tessellates the regions with galaxy signal, with the resulting bins clustering tightly around the target $\sn$. The empty, noise-dominated regions are efficiently grouped into very large bins, which a user would typically discard in a subsequent analysis step.

The algorithm's success in this regime is due to two key design choices. First, the bin-accretion stage seeds each new bin from the unbinned pixel with the highest $\sn$. This allows the algorithm to effectively grow bins `in parallel' around multiple, disconnected sources. This contrasts with the classic \textsc{VorBin} approach, which started new bins from the barycentre of all remaining pixels and could risk a large noise-bin growing to engulf a faint, distant source. Second, unlike some previous implementations of \textsc{VorBin}, the new accretion algorithm does not stop if adding a pixel temporarily decreases the bin's $\sn$. Instead, it continues to accrete neighbours until the target capacity is met or the bin's roundness criterion is violated. This makes the process more robust in the background-dominated regime.
 
This example suggests that one could apply \textsc{PowerBin} to an entire data cube and select the scientifically useful bins afterward. However, for fields with extremely sparse signals, a conservative approach of pre-selecting regions of interest before binning may still be a sensible choice.

\subsection{Application on Real IFS Data}
  
To provide a direct comparison with previous work, I apply \textsc{PowerBin} to the SAURON integral-field data of the galaxy NGC 2273, shown in \autoref{fig:powerbin_ngc2273}. This dataset served as the primary test case for the original Voronoi-binning method and its subsequent multiplicatively-weighted extension, making it a benchmark for two decades. 
The application of those previous methods to this same dataset is shown in \citet[][fig.~8]{Cappellari2003} and \citet[][fig.~5]{Diehl2006} respectively, and can be directly compared with my \autoref{fig:powerbin_ngc2273}. 
The figure shows that the new algorithm performs flawlessly on this classic dataset. It produces a clean, convex tessellation that adapts to the galaxy's morphology, and the resulting bin $\sn$ is highly uniform around the target value. The example included in the public \textsc{VorBin} package, using the WVT regularization, gives a fractional rms scatter of 7.3\%, which is slightly larger than the 6.5\% scatter produced by \textsc{PowerBin} on the same input data. This confirms that \textsc{PowerBin} successfully matches the results of the original method on real astronomical data.

This paper focuses on the algorithm and does not present new scientific applications, such as the generation of kinematic maps. This is because, while \textsc{PowerBin} resolves the critical issues of non-convex bins and slow computation, the resulting tessellations are visually very similar to those from the classic \textsc{VorBin} method. The latter has been successfully applied to hundreds of datasets over many years.

Instead of reproducing similar science results, I refer the reader to existing work for examples of the high-quality scientific products that can be derived from this binning approach. For beautiful maps of stellar kinematics and populations, particularly from high-quality MUSE data, see, for instance, \citet{Krajnovic2015, Mitzkus2017, Gadotti2019, Gadotti2020, Bittner2020}. For the largest applications of \textsc{VorBin} to date on the ever-increasing samples from major IFS surveys, see the results from the ATLAS\textsuperscript{3D}, CALIFA, SAMI, and MaNGA surveys in \citet{Cappellari2011p1}, \citet{FalconBarroso2017}, \citet{vandeSande2017}, and \citet{Westfall2019}, respectively.

\begin{figure}
  \centering
  \includegraphics[width=\columnwidth]{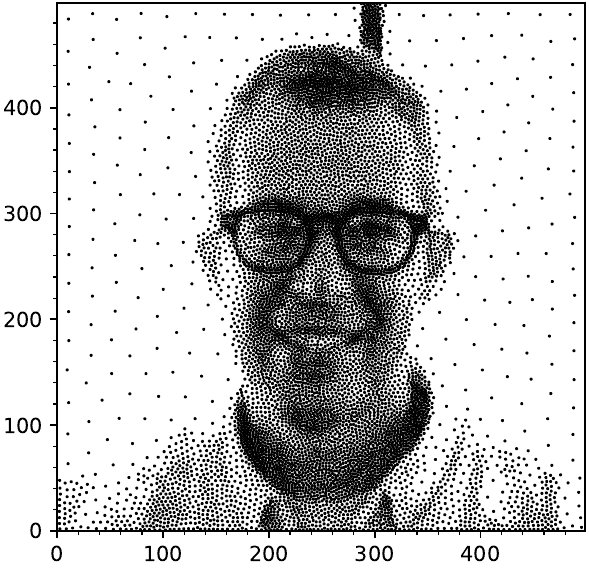}
    \caption{Demonstration of the \textsc{PowerBin} algorithm's ability to handle large datasets with sharp, irregular discontinuities. For this test, I used a $500 \times 500$ pixel greyscale self-portrait. The flux was inverted so that dark regions correspond to high signal, and the algorithm was tasked with partitioning the image into $10^4$ bins of equal integrated flux. The figure shows the final positions of the bin generators. As expected, the generators form a `blue noise' distribution, with their density tracing the underlying signal. This illustrates the connection between capacity-constrained power diagrams and stippling algorithms used in computer graphics. This example highlights the versatility and scalability of \textsc{PowerBin} for a wide range of applications beyond astronomical data.}
  \label{fig:powerbin_michele}
\end{figure}

\begin{figure}
	\includegraphics[width=\columnwidth]{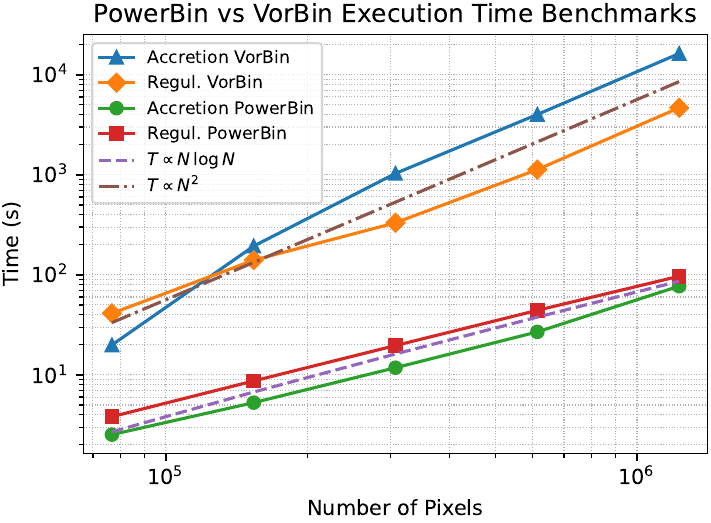}
	    \caption{Benchmark comparison of the computational time for the classic \textsc{VorBin} algorithm and the new \textsc{PowerBin} method. The plot shows the execution time as a function of the number of input pixels, $N$, for a simulated galaxy image. The four curves represent the two main stages of each algorithm: bin accretion and iterative regularization. The classic method (\textsc{VorBin}, top two curves) shows a steep scaling that approaches the theoretical $\mathcal{O}(N^2)$ complexity of multiplicatively-weighted Voronoi diagrams. The new method (\textsc{PowerBin}, bottom two curves) demonstrates a significantly improved performance, closely following the optimal $\mathcal{O}(N\log N)$ scaling expected for power diagrams. Crucially, the bin-accretion stage of \textsc{PowerBin} was also dramatically improved to follow a similar scaling.
		\label{fig:powerbin_benchmark}}
\end{figure}

\subsection{General-Purpose Tessellation}

Finally, to demonstrate the scalability and versatility of \textsc{PowerBin} on a non-astronomical problem, I apply it to a task common in computer graphics: creating a stipple drawing from an image. I took a $512 \times 512$ pixel greyscale self-portrait (\autoref{fig:powerbin_michele}) and tasked the algorithm with partitioning it into $10^4$ bins of equal integrated flux. To achieve the desired artistic effect, the image was inverted so that dark regions correspond to high signal.

The result, shown in \autoref{fig:powerbin_michele}, plots the final positions of the bin generators. The algorithm successfully handles this large and complex input, which features sharp, irregular discontinuities. As expected from the connection to optimal transport, the generators form a `blue noise' point distribution, where their density traces the underlying signal structure. This example showcases the computational efficiency and robustness of \textsc{PowerBin} on large datasets. The entire process took just 20 seconds on a standard laptop, with 12 seconds for the bin-accretion phase and 8 seconds for the CPD regularization. This highlights its potential as a general-purpose tool for capacity-constrained tessellation in a wide variety of scientific and technical fields.

\section{Execution Time Benchmarks}
\label{sec:benchmark}

To quantify the performance improvement of the new algorithm, I conducted a series of benchmark tests comparing the execution time of \textsc{PowerBin} against the classic \textsc{VorBin} package. All tests were performed on a standard laptop with an Intel i7-1355U processor. The process was running on a single core at a sustained frequency of about 3GHz. The results are presented in \autoref{fig:powerbin_benchmark}.

For these tests, I generated a sequence of mock galaxy images of increasing size. The input signal for all tests was a simulated galaxy with an exponential surface brightness profile (Sérsic $n_{\rm Ser}=1$), a fixed axial ratio of 4/3, and Poisson noise, corresponding to the uncorrelated noise case shown in the top panel of \autoref{fig:powerbin_sersic_stack}. I created five images, starting at $320\times240$ pixels and progressively doubling the total number of pixels up to $1280\times960$. For each image, the target signal-to-noise was chosen to make the number of bins, $n$, scale proportionally with the total number of pixels, $N$. This approach, which keeps the average number of pixels per bin constant, simulates a common scientific goal: exploiting a larger number of pixels to increase the spatial sampling of the binned map (i.e., more bins). This setup provides a realistic benchmark of how the algorithm's performance scales as both the number of input pixels and output bins increase. I adopted a number of bins logarithmically spaced from $n=1600$ to $n=25600$.

\autoref{fig:powerbin_benchmark} plots the execution time versus the number of input pixels on a log-log scale for the two main computational stages: the accretion and regularization steps for both the old \textsc{VorBin} and the new \textsc{PowerBin} algorithms. The performance difference is dramatic and confirms the expected theoretical scaling laws.

The classic \textsc{VorBin} method, shown by the upper two curves, exhibits a computational time that scales significantly more steeply than $\mathcal{O}(N\log N)$. At large $N$, both its accretion (blue triangles) and regularization (orange diamonds) stages approach a scaling consistent with $\mathcal{O}(N^2)$ (dot-dashed brown line). This is the expected behaviour, as the regularization is based on a MWV diagram, which has a quadratic time complexity. Similarly, the classic bin-accretion algorithm also performs operations on average linear in $N$ for every pixel, leading to a $\mathcal{O}(N^2)$ time complexity.

In stark contrast, the new \textsc{PowerBin} algorithm, shown by the lower two curves, demonstrates vastly superior performance. Both the new fast bin-accretion stage (green circles) and the Centroidal Power Diagram regularization (red squares) follow a trend that is nearly perfectly described by the theoretical $\mathcal{O}(N\log N)$ scaling (dashed purple line). This is the optimal complexity for this class of geometric problem and is a direct result of the algorithmic improvements described in \autoref{sec:fast_power_insight} and \autoref{sec:accretion}. The bottom line is that for a dataset with one million pixels, the new algorithm is approximately two orders of magnitude faster than the previous standard, turning a computation that would take 6 hours into one that takes 3 minutes. This efficiency gain is critical for the practical analysis of large-scale astronomical surveys.

It is important to emphasize that this benchmark compares the \textit{relative} performance and algorithmic scaling of the two methods. Both the classic \textsc{VorBin} and the new \textsc{PowerBin} are implemented entirely in Python, relying on standard scientific libraries like NumPy \citep{Numpy2020} and SciPy \citep{Scipy2020}. The absolute execution times could be substantially reduced by porting the computationally-intensive parts to a compiled language or specialized hardware like GPUs. However, such optimizations would not change the fundamental time complexity of the algorithms. The key result of this comparison is the difference in scaling---$\mathcal{O}(N\log N)$ versus $\mathcal{O}(N^2)$---which demonstrates the inherent efficiency of the new approach, independent of the specific implementation.

\section{Conclusions}
\label{sec:conclusions}

In this paper, I have introduced \textsc{PowerBin}, a new algorithm for the adaptive binning of two-dimensional data. This work was motivated by the increasing scale of modern astronomical surveys and the limitations of existing methods, which are either too slow or lack guarantees of bin convexity. The main contributions of this work can be summarized as follows:

\begin{enumerate}
    \item \textbf{A New Theoretical Framework:} I have framed the adaptive binning problem within the mathematical theory of optimal transport. The natural solution in this framework is a Centroidal Power Diagram (CPD), a generalization of a Centroidal Voronoi Tessellation that rigorously accommodates capacity constraints while guaranteeing convex bins.

    \item \textbf{A Fast and Robust Heuristic Solver:} Formal CPD solvers, based on gradient-based optimization of a dual energy functional, are ill-suited to real astronomical data, which is discrete and often has non-additive noise properties. I have introduced a novel heuristic algorithm that circumvents these issues. It is based on a simple physical insight into the geometry of packed cells, which provides a direct, non-linear update rule for the power diagram weights to enforce the target capacity. This approach is fast, robust, and converges reliably even when the capacity function is non-additive, a regime where formal methods fail.

    \item \textbf{An Optimized Bin-Accretion Algorithm:} The bin-accretion stage, which provides the crucial starting point for the iterative refinement, would have become the computational bottleneck of the current methods. Therefore, I developed a new implementation with near-linear time complexity, $\mathcal{O}(N\log N)$. This is achieved by using a pre-computed Delaunay triangulation for adjacency information, a frontier-based growth strategy, incremental updates for bin properties, and a heap-managed frontier to efficiently select pixels.

    \item \textbf{Superior Performance and Scalability:} The combination of the fast CPD solver and the optimized bin-accretion algorithm results in a dramatic performance improvement. Benchmark tests show that the entire \textsc{PowerBin} algorithm scales as $\mathcal{O}(N\log N)$, in stark contrast to the $\mathcal{O}(N^2)$ scaling of previous methods. For a dataset of one million pixels, the new algorithm is approximately two orders of magnitude faster than the widely-used \textsc{VorBin} package.
\end{enumerate}

I have demonstrated through a series of tests on simulated and real data that \textsc{PowerBin} produces high-quality, convex tessellations with excellent capacity uniformity. It successfully handles a wide range of signal distributions and is robust to the challenges of correlated noise. Its robustness was further demonstrated on challenging background-subtracted mock observations containing multiple objects, where the noise is highly non-Poissonian. Its performance on the classic SAURON data of NGC 2273 confirms that it reproduces and improves upon the results of the original Voronoi-binning method.

By addressing the key limitations of speed and convexity, \textsc{PowerBin} provides a powerful and scalable tool for the analysis of the massive datasets generated by current and future astronomical surveys. Its versatility, demonstrated on a non-astronomical imaging problem, also suggests its potential for broad application in other scientific and technical fields requiring capacity-constrained tessellation. Notably, both the bin-accretion and regularization stages of \textsc{PowerBin} extend naturally to higher dimensions with minimal code changes and no change in the underlying concept. The Python implementation of the algorithm is publicly available to the community in the \textsc{PowerBin} package\footnote{\url{https://pypi.org/project/powerbin/}}.

\section*{Acknowledgements}

I thank the referee for a constructive and insightful report that helped improve the clarity and completeness of this paper.
I am grateful for the serene setting of the island of Pellestrina, whose tranquil location between the Venetian Lagoon and the Adriatic Sea proved to be the perfect environment for the writing of this paper. A special thanks to my family, who demonstrated saintly patience while enduring countless updates about the \textsc{PowerBin} algorithm.

\section*{Data Availability}

No new data were generated in support of this research. The \textsc{PowerBin} package is available at \url{https://pypi.org/project/powerbin/}.

\bsp    
\label{lastpage}
\end{document}